\documentclass{moriond}

\def\be{\begin{equation}}
\def\ee{\end{equation}}
\def\bea{\begin{eqnarray}}
\def\eea{\end{eqnarray}}

\newcommand{\Photo}{\includegraphics[height=35mm]{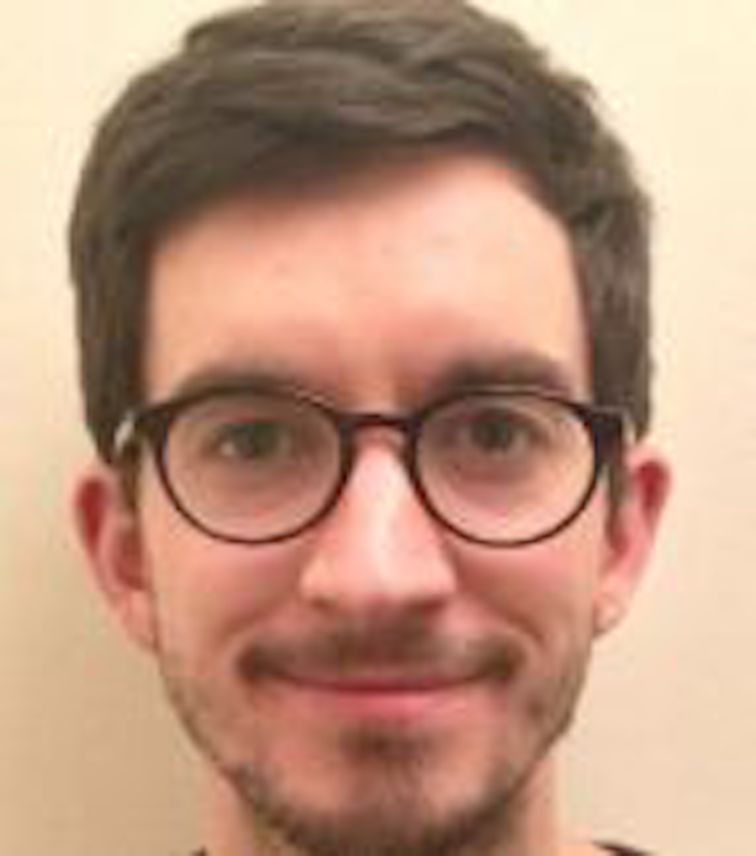}}

\begin{document}
\vspace*{4cm}
\title{Physics from Photons at the LHC}

\author{L. A. Harland--Lang}

\address{Rudolf Peierls Centre, Beecroft Building, Parks Road, Oxford, OX1 3PU \vspace*{0.5cm}}

\maketitle\abstracts{LHC collisions can act as a source of photons in the initial state. This mechanism plays an important role in the production of particles with electroweak couplings, and a precise account of photon-initiated (PI) production at the LHC is a key ingredient in the LHC precision physics programme. I will  discuss the possibility of modelling PI processes directly via the structure function approach. This can provide percent level precision in the  production cross sections, and is therefore well positioned to account for LHC precision requirements. This formalism in addition allows one to make use of another useful feature of photons, namely that they are colour-singlet and can often be emitted elastically (or quasi-elastically) from the proton. I will discuss recent work on applications of the structure function approach to precision calculations of PI production in the inclusive mode, and to 'exclusive' processes with rapidity gaps, which can provide a unique probe of the Standard Model and physics beyond it.}

\section{The Structure Function Approach and inclusive photon--initiated production}

A major aim of the LHC, and the HL--LHC upgrade that will follow, is to precisely test the Standard Model (SM) predictions for as wide a range of collider processes as possible. A particularly important element of this involves events with leptons in the final state, which play a key role in determinations of the weak mixing angle, $\sin^2 \theta_W$, the $W$ boson mass, $M_W$, and  constraints on the proton PDFs.

A key ingredient in this  is the availability of high precision theoretical predictions for the SM  processes, an important element of which is the contribution from photon--initiated (PI) channel. A rather useful method~\cite{Harland-Lang:2019eai,Harland-Lang:2021zvr} to provide this is to work in the `structure function' (SF) framework. Here, one calculates the cross section directly in terms of the proton structure functions. For processes such as (off $Z$--peak) lepton pair production this provides percent precision in the predicted cross sections, with no accompanying factorization scale variation uncertainty, as is present in the calculation within collinear factorization.

 \begin{figure}[h]
\begin{center}
\includegraphics[scale=0.65]{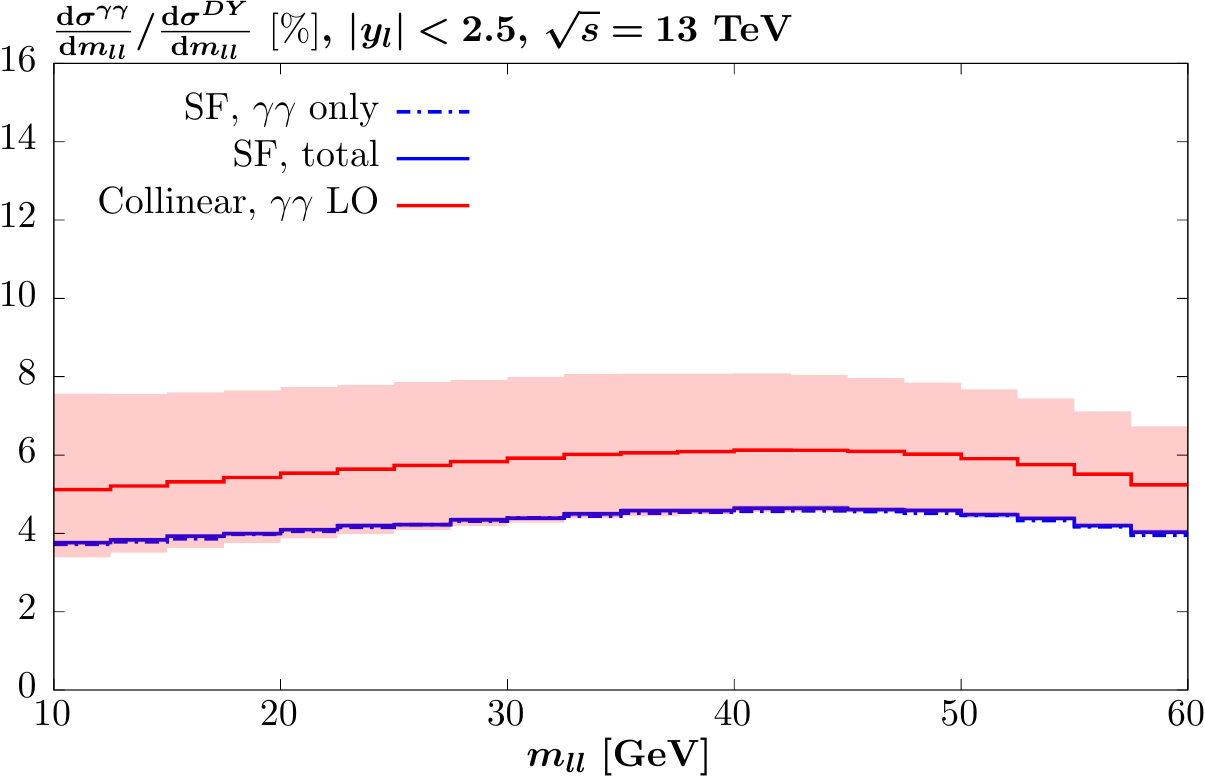}
\includegraphics[scale=0.65]{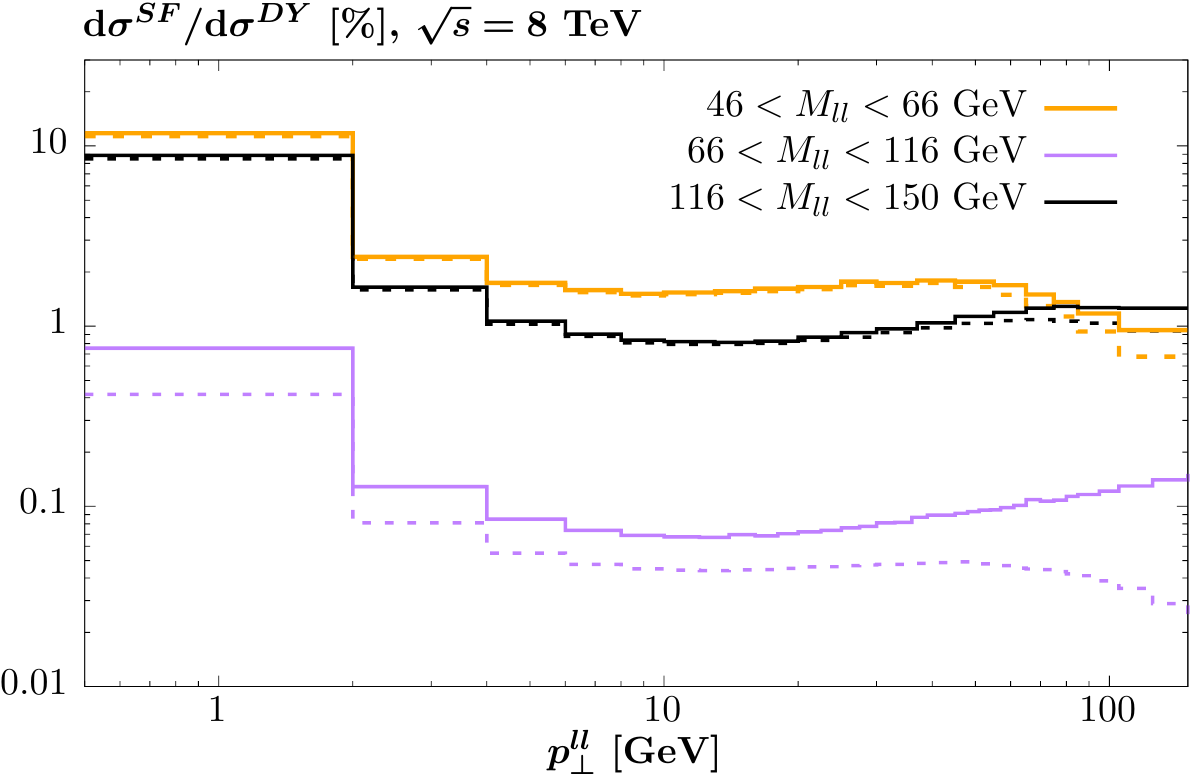}
\caption{\small{(Left) Ratio of the PI cross sections for lepton pair production to the NLO QCD Drell--Yan cross section at the $13$ TeV LHC.  The LO collinear predictions (including scale variation uncertainties) and the structure function result are shown, in the latter case for both the pure $\gamma\gamma$ initiated and the total. (Right) Percentage contribution from PI production to the lepton pair $p_\perp$ distribution, within the ATLAS~\protect\cite{Aad:2015auj} off--peak event selection, at 8 TeV. The QCD predictions correspond to NNLO + NNLL QCD theory~\protect\cite{Bizon:2018foh}. The total (pure $\gamma\gamma$) contributions are shown by the solid (dashed) lines.}}\label{fig:mDY}
\end{center}
\end{figure}

In the SF approach, the PI cross section is given by
  \be\label{eq:sighh}
  \sigma_{pp} = \frac{1}{2s}  \int \frac{{\rm d}^3 p_1 {\rm d}^3 p_2 {\rm d}\Gamma}{E_1 E_2}  \alpha(Q_1^2)\alpha(Q_2^2)
  \frac{\rho_1^{\mu\mu'}\rho_2^{\nu\nu'} M^*_{\mu'\nu'}M_{\mu\nu}}{q_1^2q_2^2}\delta^{(4)}(q_1+q_2 - k)\;.
 \ee
 Here the outgoing hadronic systems have momenta $p_{1,2}$ and the photons have momenta $q_{1,2}$, with $q_{1,2}^2 = -Q_{1,2}^2$. We consider the production of a system of 4--momentum $k = q_1 + q_2 = \sum_{j=1}^N k_j$ of $N$ particles, where ${\rm d}\Gamma = \prod_{j=1}^N {\rm d}^3 k_j / 2 E_j (2\pi)^3$ is the  phase space volume. $M^{\mu\nu}$ corresponds to the $\gamma\gamma \to X(k)$ production amplitude, with arbitrary photon virtualities. $\rho$ is the density matrix of the virtual photon, which is given in terms of the well known proton structure functions, see~\cite{Harland-Lang:2019eai,Harland-Lang:2021zvr} for an explicit expression and~\cite{Harland-Lang:2021zvr} for the extension to include initial--state $Z$ bosons.
 
 A representative selection of results are shown in Fig.~\ref{fig:mDY}. In the left plot we show the lepton pair invariant mass distribution, plotting the ratio of the PI contribution to the NLO QCD Drell--Yan cross section at the $13$ TeV LHC. For this choice of cuts, the PI component is at the $\sim 4\%$ level, and hence is small but certainly not negligible. We note that the solid curve includes the uncertainty due to the experimental determination of the structure functions, but this is so small as to be not visible on the plot. The `total' contribution includes initial--state $Z$ boson, and mixed $\gamma/Z + q$ contributions, but these are found to give a negligible contribution. The LO collinear result is also shown for comparison. This is seen to lie above the SF results, though consistent within the large scale variation uncertainties; clearly, one should work at least at NLO when applying the collinear approach. However, for the present observable we can see that the SF approach provides percent level precision already. 
 
 Above the $Z$ peak (not shown here), the PI contribution is as large as $\sim 10\%$, and can again be predicted with high precision by the SF approach. As discussed in further detail in~\cite{Harland-Lang:2021zvr}, these results imply that the PI contribution to the dilepton $\cos\theta^*$ distribution, which is used for determinations of $\sin^2 \theta_W$ as well as PDF constraints, can also be highly relevant. Detailed comparisons are presented in~\cite{Harland-Lang:2021zvr} and this is indeed found to be the case, with the SF approach providing high precision predictions for the PI contribution.
 
As the SF formulation of Eq.~\ref{eq:sighh} is provided differentially with respect to the photon virtualities, $Q^2$, we can also readily provide predictions with respect to the dilepton transverse momentum, $p_\perp^{ll}$. In Fig.~\ref{fig:mDY} (right) we show the ratio of the PI contribution to NNLO+N${}^3$LL resummed QCD predictions produced with \texttt{NNLOjet+RadISH}~\cite{Bizon:2018foh}. The total (solid) curve includes mixed  $\gamma + q$ diagrams, which will be sensitive to resummation effects in the low $p_\perp^{ll}$ region, not included here,  is only shown for rough guidance in this region. Focussing on the pure $\gamma\gamma$ component, i.e. the dashed lines, one can see that is a significant enhancement observed in the pure $\gamma\gamma$ contribution in the lower $0 < p_\perp^{ll} < 2$ GeV region. This is explained in part by the Sudakov suppression in the QCD contribution in this region, which is absent in the $\gamma\gamma$ channel. However, another key factor in this is that the $\gamma\gamma$ cross section is particularly peaked in this region, due to the significant contribution from elastic photon emission. This elastic component, or indeed the low $Q^2$ resonant and non--resonant components, are not modelled differentially in a pure collinear calculation, and hence the SF calculation is particularly well suited to deal with this very  low $p_\perp^{ll}$ region. As explored in~\cite{Harland-Lang:2021zvr}, this could have implications for experimental determinations of $M_W$, through the tuning that is done to the low $p_\perp^{ll}$ region of dilepton production. 

Finally, we note that the benefit of applying the SF approach directly, while transparent for process where the final state of interest ($l^+ l^-$, $W^+ W^-$...) is directly produced by the $\gamma\gamma$ initial state, is less clear in the mixed $\gamma + q$ case. Here, one must deal with the collinear enhancement of the $\gamma \to q\overline{q}$ splitting, and at this level of precision include QED DGLAP evolution of the quark/antiquark PDFs, which certainly requires one introduce a photon PDF within the LUXqed approach. Further discussion can be found in~\cite{Harland-Lang:2021zvr}.

\section{Modelling (semi)--exclusive photon--initiated production}
 
 A  feature of the PI channel in proton--proton collisions is that the colour singlet photon exchange naturally leads to exclusive events, where the photons are emitted elastically from the protons. This is particularly relevant in the context of the dedicated forward proton detectors at the LHC, namely  AFP and CT--PPS, which have been installed in association with both ATLAS and CMS, respectively~\cite{Royon:2015tfa}. More generally, even if the initial--state photon is emitted inelastically, there is no colour flow as a result, and there is still a possibility for semi--exclusive events with rapidity gaps  in the final--state between the proton dissociation system(s) and the centrally produced object. Indeed, a range of data  have been collected using this technique at the LHC~\cite{Harland-Lang:2020veo}.
 
 This theoretical treatment of this class of events is rather distinct from the standard inclusive case. The reasons for this are twofold: first, events where decay products from the proton dissociation system enter the veto region must be excluded, and second, there may be additional inelastic proton--proton QCD interactions (in other words, underlying event activity) that fill the gap region. The latter effect must be accounted for via the so--called `survival factor' probability of no additional proton--proton interactions~\cite{Harland-Lang:2014lxa}, while the former requires a fully differential treatment of the PI process, including a MC implementation such that the showering and hadronisation of the dissociation system may be accounted for.

 In~\cite{Harland-Lang:2020veo}, we presented such a MC implementation, \texttt{SuperChic 4}, for the case of lepton pair production. This makes use of the SF approach, to provide a high precision prediction for the underlying PI process that is fully differential in the kinematics of the final--state protons and/or dissociation systems. This can then be interfaced to a general purpose MC for further showering/hadronization; we make use of \texttt{Pythia 8.2}~\cite{Sjostrand:2014zea}. We in addition account for the survival factor, in a manner that take full account of the dependence of this quantity on the event kinematics and the specific channel (elastic or inelastic). \texttt{SuperChic 4} is the first generator of its kind to take account of all of these features, which are essential when providing results for semi--exclusive PI production. 
 
\begin{figure}
\begin{center}
\includegraphics[scale=0.63]{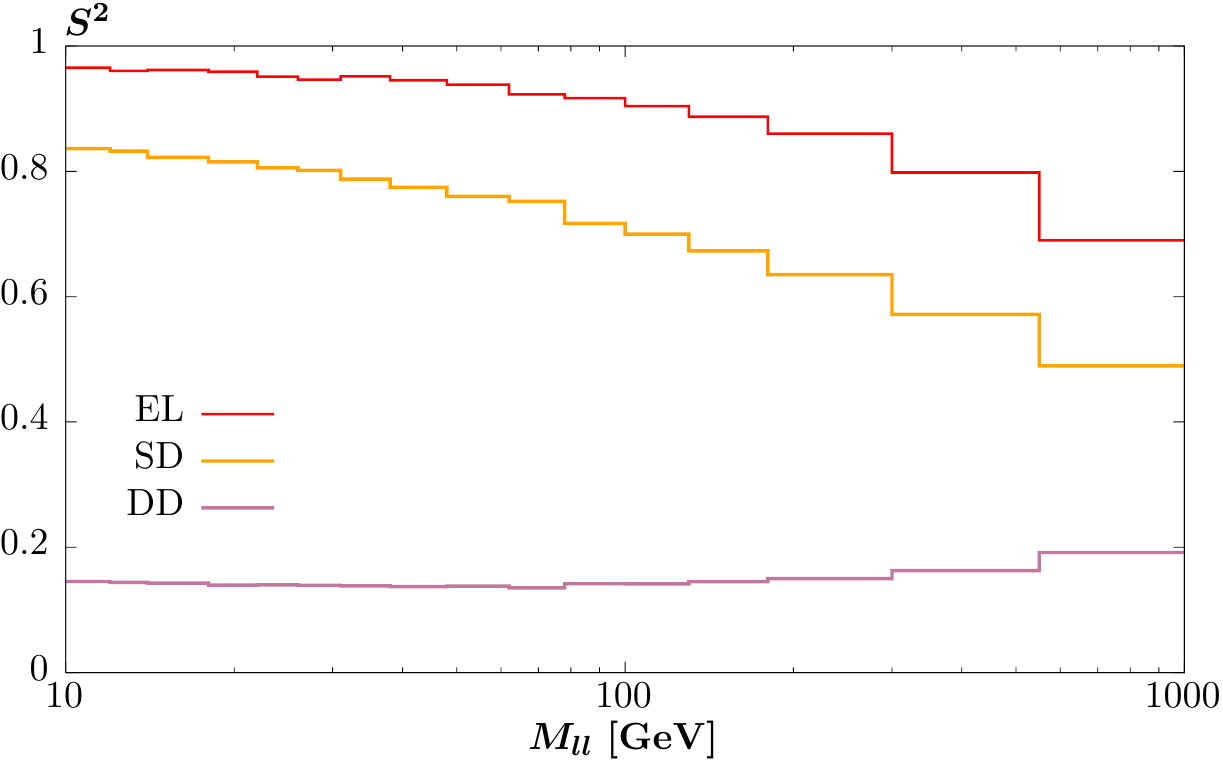}
\includegraphics[scale=0.38]{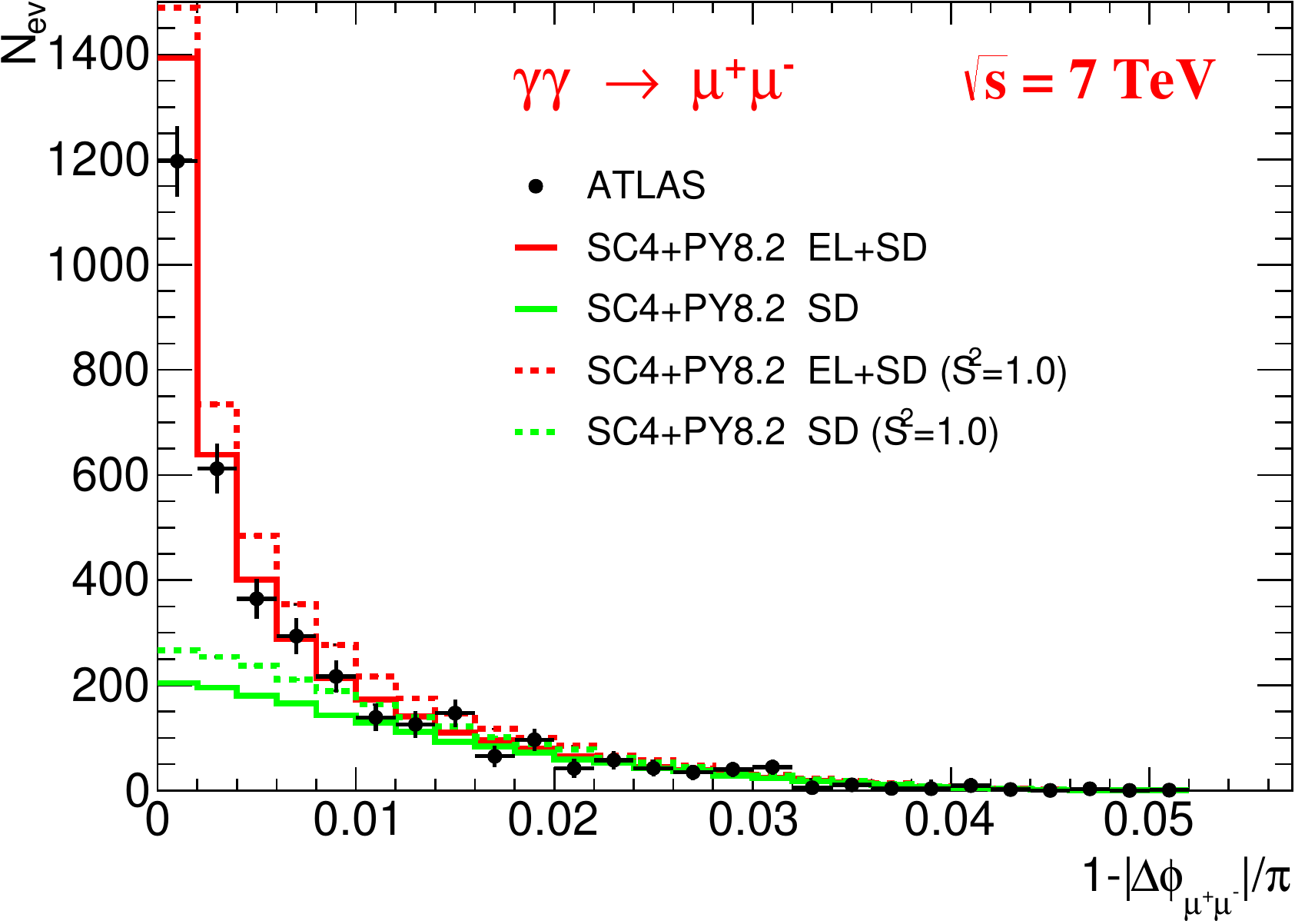}
\caption{\small{(Left) Soft survival factor for lepton pair production as a function of the invariant mass, $M_{ll}$, of the dilepton system. (Right) Comparison of \texttt{SuperChic 4 + Pythia 8.2} predictions for the dilepton acoplanarity distribution compared to the ATLAS data~\protect\cite{Aad:2015bwa} at $\sqrt{s}=7$ TeV, within the corresponding experimental fiducial region, and with a rapidity veto applied on tracks in the central region. Results without the soft survival factor included are shown in addition.}}\label{fig:S2}
\end{center}
\end{figure}

In Fig.~\ref{fig:S2} (left) we show the predicted survival factor as a function of the dimuon invariant mass, at $\sqrt{s}=13$ TeV. We can see that broadly there is a large difference in the magnitude of the survival factor between the DD and elastic/SD cases, with the former being significantly smaller. This is driven by the fact that in the DD case the photon $Q^2$ is generally much higher, and so the collision is less peripheral in terms of the impact parameter of the colliding protons; the most peripheral elastic interaction has the highest survival factor. We can also see that as the invariant mass increases, the survival factor decreases, due to effect of the kinematic requirement for producing an on-shell proton at the elastic vertex for larger photon momentum fractions, which implies a larger photon $Q^2$. For the DD case the survival instead increases somewhat, due to the smaller phase space in photon $Q^2$ at the highest $M_{ll}$ values. These are examples of a broader result, namely that the survival factor is not a single constant value, but rather depends sensitively on the process and kinematics.

In Fig.~\ref{fig:S2} (right) we show a comparison of the predicted acoplanarity distribution for muon pairs to the ATLAS data on semi--exclusive dilepton production at $\sqrt{s}=7$ TeV~\cite{Aad:2015bwa}. The theory result includes the impact of the rapidity veto that is applied in order to selected these events, as well as the survival factor (we show results without this included for comparison). We can see that the distribution is described rather well once the survival factor is included.

In summary, PI production is rather unique channel at the LHC that plays a key role in both inclusive and exclusive particle production. We have presented state--of--the--art results for dilepton production in both of these channels, including MC implementations; \texttt{SFGen}~\cite{Harland-Lang:2021zvr} and  \texttt{SuperChic 4}~\cite{Harland-Lang:2020veo} for inclusive and exclusive production, respectively.

\section*{Acknowledgments}

I thank the Science and Technology Facilities Council (STFC) for support via ST/L000377/1.

\end{document}